\documentclass{article}

\PassOptionsToPackage{numbers, compress}{natbib}

\usepackage[final]{neurips_2022}
\usepackage{subfigure}
\usepackage{wrapfig}
\usepackage{lipsum} 
\usepackage{floatflt}

\usepackage[utf8]{inputenc} 
\usepackage[T1]{fontenc}    
\usepackage{hyperref}       
\usepackage{url}            
\usepackage{booktabs}       
\usepackage{amsfonts}       
\usepackage{nicefrac}       
\usepackage{microtype}      
\usepackage{xcolor}         
\usepackage{graphicx}
\usepackage{lipsum}
\usepackage{longtable}
\usepackage{array}
\usepackage{amsmath}
\usepackage{booktabs}
\usepackage{makecell}
\title{ProtoEEGNet: An Interpretable Approach for Detecting Interictal Epileptiform
Discharges}

\author{%
    Dennis Tang\thanks{Corresponding author: \texttt{dennis.tang@duke.edu}} \hspace{0.5mm}\thanks{Co-first authors} \hspace{0.5mm}$^1$, 
    Frank Willard$^\dagger$$^1$,
    Ronan Tegerdine$^1$,
    Luke Triplett$^1$,
    Jon Donnelly$^1$,
   Luke Moffett$^1$,\\
   \textbf{Lesia Semenova$^1$,
   Alina Jade Barnett$^1$,
   Jin Jing$^2$,
  Cynthia Rudin\thanks{Co-senior authors} \hspace{0.5mm}$^1$, 
  Brandon Westover$^\ddagger$$^{2 3}$}\\
  $^1$Department of Computer Science, Duke University\\
  $^2$Department of Neurology, Massachusetts General Hospital\\
  $^3$Harvard Medical School, Beth Israel Deaconess Medical Center, McCance Center for Brain Health
  }

\begin{document}

\maketitle

\begin{abstract}
In electroencephalogram (EEG) recordings, the presence of interictal epileptiform discharges (IEDs) serves as a critical biomarker for seizures or seizure-like events. Detecting IEDs can be difficult; even highly trained experts disagree on the same sample. As a result, specialists have turned to machine-learning models for assistance. However, many existing models are black boxes and do not provide any human-interpretable reasoning for their decisions. In high-stakes medical applications, it is critical to have interpretable models so that experts can validate the reasoning of the model before making important diagnoses. We introduce ProtoEEGNet, a model that achieves state-of-the-art accuracy for IED detection while additionally providing an interpretable justification for its classifications. Specifically, it can reason that one EEG looks similar to another ``prototypical'' EEG that is known to contain an IED. ProtoEEGNet can therefore help medical professionals effectively detect IEDs while maintaining a transparent decision-making process.
\end{abstract}

\section{Introduction}
Accurately identifying the presence of interictal epileptiform discharges (IEDs) is important when analyzing EEG signals, as they can be a critical biomarker for epilepsy \cite{SEIDEL2016102, ESSENTIAL_CLINICAL, VALUE_OF_EEG}. However, detecting IEDs is a challenging task even for trained experts, who frequently disagree on whether an EEG contains an IED as shown in \cite{Jing2022InterraterRO} and Figure \ref{fig:expert_disagreement} in the Appendix. This difficulty has led clinicians and researchers alike to turn towards machine learning models to assist them \cite{Jing2020DevelopmentOE, SpikeViaCNN, ScalpEEG, Raghu2020EEGBM, Tveit2023AutomatedIO}. However, many of these approaches are black boxes that assert a prediction and offer little to no explanation for their decisions. Moreover, while black box models may perform well on experimental benchmarks, they may struggle to generalize to real-world conditions \cite{Beede2020AHE, Zech2018VariableGP}. These shortcomings are especially problematic in high-stakes medical settings, as black box models can confuse or frustrate clinicians with incorrect predictions from faulty reasoning. While post-hoc analysis methods have been applied to EEG models \cite{Lopes2023UsingCS}, they were criticized for being not faithful to the original model and insufficient to ascertain the model's underlying behavior \cite{Rudin2018StopEB, adebayo2018sanity}. Therefore, it is critical to adopt inherently interpretable models in high-stakes decision-making \cite{Rudin2018StopEB, Arun2021AssessingTT} as they enable users to review the reasoning of the model for logical consistency and plausibility. 
However, existing interpretable 
methods for IED detection such as K-nearest neighbors and template matching are slower \cite{Rudin2021InterpretableML} or less accurate \cite{Nascimento2022AQA} when compared to black boxes, making them impractical for clinical use.

\begin{figure}[htp]
    \centering
    \vspace*{-4mm}
    \includegraphics[width=1\linewidth]{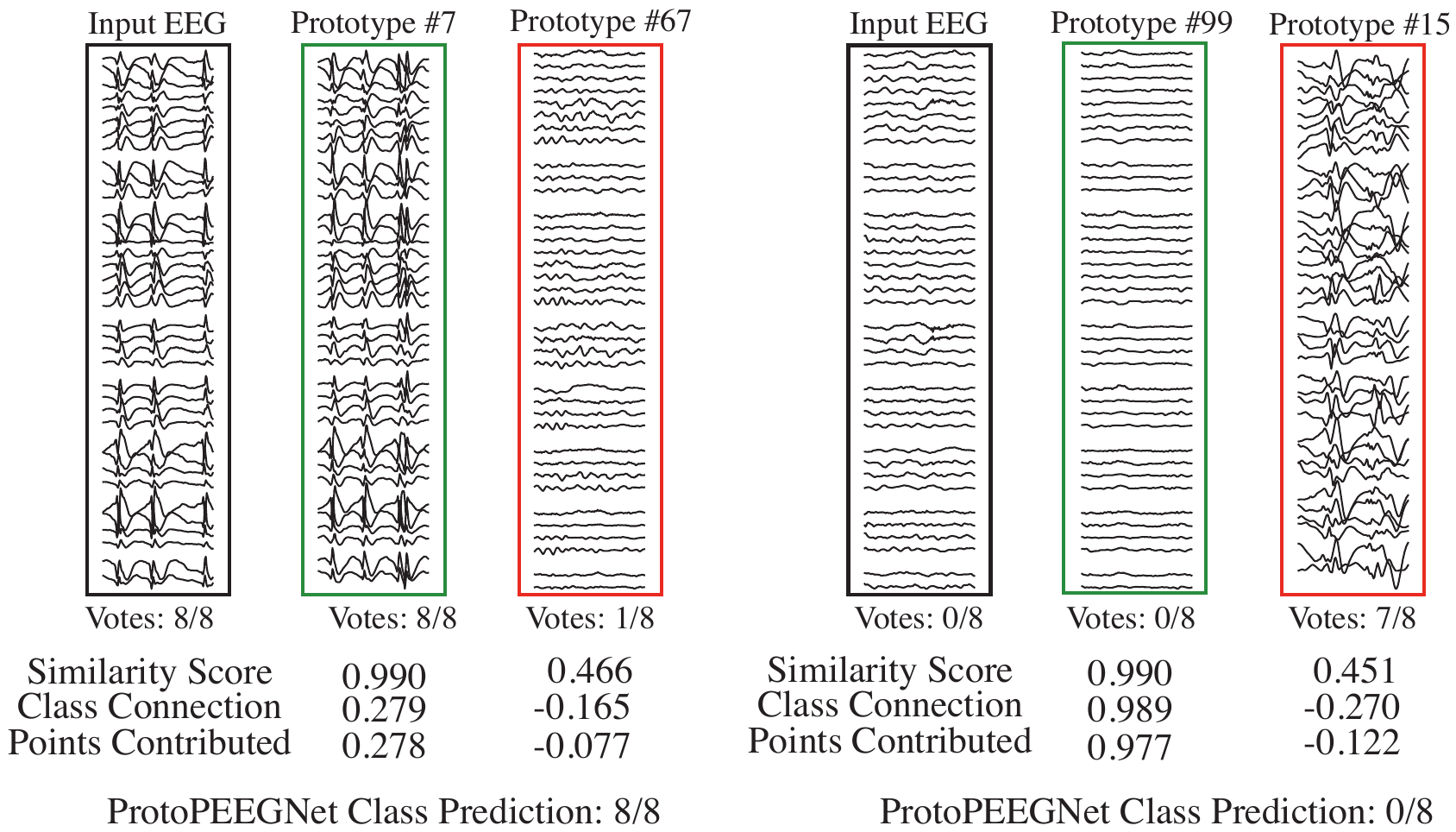}
    \label{fig:input_figure_v2}
    \vspace*{-4mm}
    \caption{\textbf{Example Of ProtoEEGNet's Reasoning.} ProtoPEEGNet makes its predictions by comparing the current EEG to learned prototypical EEGs for both spike and non-spike cases. The similarity score is calculated based on cosine similarity in the latent space representation of the EEGs. The class connection indicates the weight between the prototype and the predicted class in the final layer. The two scores are multiplied into a final ``points contributed'' score used for classification.}
\end{figure}

To address these challenges, we introduce ProtoEEGNet, an interpretable neural network that detects the occurrence of IEDs in EEGs. We demonstrate that ProtoEEGNet rivals the state-of-the-art IED detection model SpikeNet
while additionally providing a human interpretable justification for its decisions. Specifically, ProtoEEGNet learns ``prototypical'' IEDs and interprets EEGs based on their resemblance to the learned prototypes by reasoning that "This EEG looks like that EEG" (as demonstrated in Figure \ref{fig:input_figure_v2}). This aids medical experts in better understanding why our model classified a given EEG as containing an IED.

\section{Methods}

\textbf{Data.} We obtained 16,646 1-second EEG samples that spanned a range of ages and settings from patients at Massachusetts General Hospital (MGH). The EEGs were filtered (60-Hz notch, 0.5-Hz high-pass) and re-sampled to 128 Hz for consistency. Each EEG was labeled by 8 physician subspecialists who annotated whether or not they believed a sample contained an IED, following the annotation procedure of \citet{Jing2020DevelopmentOE}. Given the task difficulty and high annotator disagreement, we treated each annotator's label as a ``vote''. The proportion of expert votes indicating an EEG (0/8 to 8/8) 
was used as training labels to create 9 classes, each representing a possible number of votes. A visualization of class differences is shown in Appendix \ref{ClassExamples}. Data was partitioned into 73\% training, 12\% validation, and 15\% testing splits.

\begin{figure}[htp]
    \centering
    \includegraphics[width=1\linewidth]{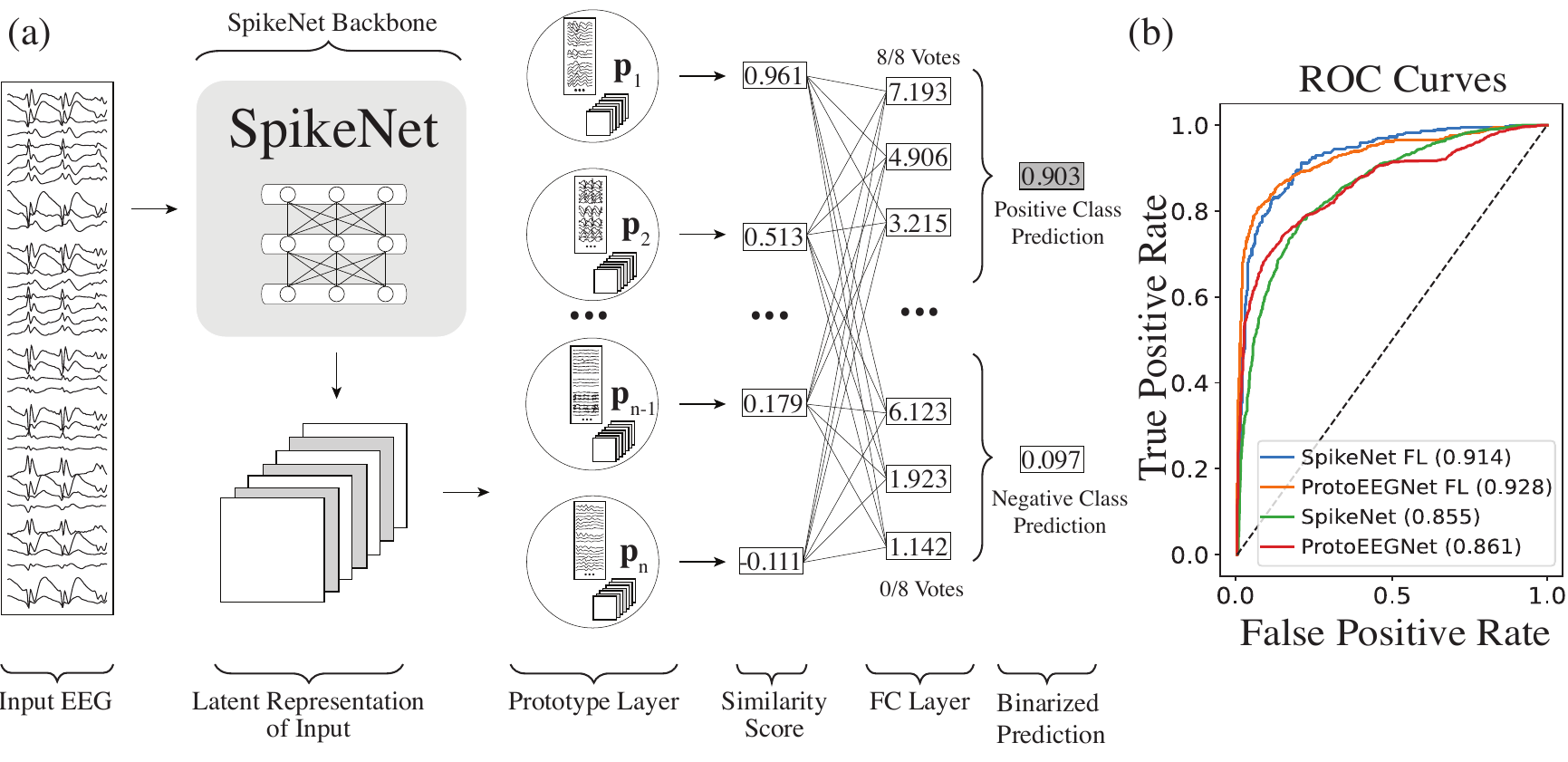}
    \vspace*{-7mm}
     \caption{ (a) \textbf{ProtoEEGNet Architecture}. An input EEG is propagated through the backbone which outputs a latent representation of the EEG. The prototype layers then compute the cosine similarities between the latent representation and the prototypes which are passed to a fully connected layer that computes classification logits. Optionally, predicted probabilities can be binarized into positive and negative class predictions when yes-no spike detection is sufficient. (b) \textbf{ROC curves for SpikeNet and ProtoEEGNet.} ``FL'' indicates that the test data excluded samples with 3/8, 4/8 and 5/8 votes. The parentheses in the legend indicate AUC values (confidence intervals reported in \ref{AUROCCI}).}
    \label{fig:interpnn}

\end{figure}

\textbf{ProtoEEGNet.} The architecture of ProtoEEGNet (shown in Figure \ref{fig:interpnn}a) builds upon that of ProtoPNet, an interpretable 
classifier composed of a pre-trained image-recognition backbone and subsequent ``prototype'' layers \cite{Chen2018ThisLL, deformable_protopnet}. As compared to ProtoPNet, ProtoEEGNet uses (1) a modified SpikeNet backbone and (2) multi-channel 1D EEG signals rather than 2D images. By learning to identify 9 classes based on the proportion of expert votes instead of binary labels, ProtoEEGNet is able to learn prototypes associated with multiple levels of ambiguity (e.g., 8/8 vote versus 4/8 vote prototypes).
At inference time, the latent representation of a new sample is compared with the latent representation of all learned prototypes to compute a vector of cosine similarity scores, as detailed in Appendix \ref{DefinitionAppendix} and \ref{TrainingAppendix}. These scores are then passed through a linear layer that computes class logits.

For the backbone, we adopted SpikeNet \cite{Jing2020DevelopmentOE}, an uninterpretable CNN for IED detection. SpikeNet was modified by: (1) replacing BatchNorm with LayerNorm layers due to their empirical improvement and (2) increasing the kernel size of the final convolution layer from 8 to 10. This increase produced a 1x1 latent output, ensuring that ProtoEEGNet prototypes represented an entire EEG (as in Figure \ref{fig:interpnn}) rather than a specific sub-feature. EEGs have nuanced, inter-channel dependencies such that it is necessary to evaluate EEGs against full montage prototypes \cite{Kane2017ARG}. We also trained the model end-to-end, thereby updating the weights of the SpikeNet backbone to fit the new architecture. A detailed description of the model can be found in Appendix \ref{DefinitionAppendix}.

At inference time, the latent representation of a new sample is compared with the latent representation of all learned prototypes to compute a vector of cosine similarity scores. These scores are then passed through a linear layer that computes class logits.

\textbf{Training} ProtoEEGNet is trained end-to-end for 130 epochs and uses the Adam optimizer. The model was trained to find 12 prototypes per class for a total of 108 prototypes.
We provide a more formal description of ProtoEEGNet and more details on the training procedure in Appendix \ref{DefinitionAppendix}.

\section{Results}

Since SpikeNet achieved state-of-the-art IED detection accuracy on an MGH dataset of 1051 EEGs \cite{Jing2020DevelopmentOE}, we used it as our baseline. Since our data is a superset of the SpikeNet dataset, we ensured that no samples used to train SpikeNet were partitioned into our test set. 
To compare ProtoEEGNet against SpikeNet, we binarized our model's 9-class predictions by averaging the positive class (4 votes or more) and negative class (3 votes or less) prediction probabilities and taking a softmax over the two-class probabilities. Then, given the binary prediction, we calculated the AUROC of both models on the held-out test set.
Originally in \citet{Jing2020DevelopmentOE}, the ROC was calculated for SpikeNet by filtering out samples with 3/8, 4/8, or 5/8 votes. We follow this schema to create a filtered dataset and report results for both filtered and unfiltered test data in Figure \ref{fig:interpnn}b.


We find that on both the filtered and the unfiltered data ProtoEEGNet (filtered AUC = 0.928, unfiltered AUC = 0.861) slightly outperforms SpikeNet (filtered AUC = 0.914, unfiltered AUC = 0.855).
Further, ProtoEEGNet follows interpretable reasoning as demonstrated in Figure \ref{fig:input_figure_v2}. This is a major practical advantage over SpikeNet, as practitioners may ``fact check'' model reasoning.

\section{Conclusion}
We present ProtoEEGNet, an interpretable method for detecting IEDs that provides its reasoning on a case-by-case basis. One limitation of ProtoEEGNet is ensuring each sample is well-represented by at least one prototype. We leave this challenge of increasing prototype diversity for future work. In this work, we demonstrated that the ProtoEEGNet outperforms SpikeNet (current state-of-the-art) at IED detection while
allowing users to visually verify its predictions.

\section{Potential Negative Societal Impacts}

Similar to any other classification model, ProtoEEGNet is not guaranteed to have perfect accuracy on new samples. Therefore, over-reliance on the model may result in confusion or even incorrect diagnosis in certain instances. Additionally, the model may be prone to adversarial privacy attacks which may allow attackers to recreate training samples or generate inferences about the patient training data. An exploration of the model's resistance to privacy attacks is beyond the scope of this work.

\section{Acknowledgement}

We acknowledge funding from the National Science Foundation under grant HRD-2222336.

\bibliographystyle{unsrtnat}

\bibliography{reference.bib}

\begin{thebibliography}{20}
\providecommand{\natexlab}[1]{#1}
\providecommand{\url}[1]{\texttt{#1}}
\expandafter\ifx\csname urlstyle\endcsname\relax
  \providecommand{\doi}[1]{doi: #1}\else
  \providecommand{\doi}{doi: \begingroup \urlstyle{rm}\Url}\fi

\bibitem[Seidel et~al.(2016)Seidel, Pablik, Aull-Watschinger, Seidl, and
  Pataraia]{SEIDEL2016102}
Stefan Seidel, Eleonore Pablik, Susanne Aull-Watschinger, Birgit Seidl, and
  Ekaterina Pataraia.
\newblock Incidental epileptiform discharges in patients of a tertiary centre.
\newblock \emph{Clinical Neurophysiology}, 127\penalty0 (1):\penalty0 102--107,
  2016.
\newblock ISSN 1388-2457.
\newblock \doi{https://doi.org/10.1016/j.clinph.2015.02.056}.
\newblock URL
  \url{https://www.sciencedirect.com/science/article/pii/S1388245715001613}.

\bibitem[Fountain and Freeman(2006)]{ESSENTIAL_CLINICAL}
Nathan~B. Fountain and John~M. Freeman.
\newblock Eeg is an essential clinical tool: Pro and con.
\newblock \emph{Epilepsia}, 47\penalty0 (s1):\penalty0 23--25, 2006.
\newblock \doi{https://doi.org/10.1111/j.1528-1167.2006.00655.x}.
\newblock URL
  \url{https://onlinelibrary.wiley.com/doi/abs/10.1111/j.1528-1167.2006.00655.x}.

\bibitem[van Donselaar et~al.(1992)van Donselaar, Schimsheimer, Geerts, and
  Declerck]{VALUE_OF_EEG}
Cees~A. van Donselaar, Robert-Jan Schimsheimer, Ada~T. Geerts, and August~C.
  Declerck.
\newblock {Value of the Electroencephalogram in Adult Patients With Untreated
  Idiopathic First Seizures}.
\newblock \emph{Archives of Neurology}, 49\penalty0 (3):\penalty0 231--237, 03
  1992.
\newblock ISSN 0003-9942.
\newblock \doi{10.1001/archneur.1992.00530270045017}.
\newblock URL \url{https://doi.org/10.1001/archneur.1992.00530270045017}.

\bibitem[Jing et~al.(2022)Jing, Ge, Struck, Fernandes, linda Qiao, An, Fatima,
  Herlopian, Karakis, Halford, Ng, Johnson, Appavu, Sarkis, Osman, Kaplan,
  Dhakar, Jayagopal, Sheikh, Taraschenko, Schmitt, Haider, Kim, Swisher,
  Gaspard, Cervenka, Ruiz, Lee, Tabaeizadeh, Gilmore, Nordstrom, Yoo, Holmes,
  Herman, Williams, Pathmanathan, Nascimento, Fan, Nasiri, Shafi, Cash, Hoch,
  Cole, Rosenthal, Zafar, Sun, and Westover]{Jing2022InterraterRO}
Jin Jing, Wendong Ge, Aaron~F. Struck, Marta Fernandes, linda Qiao, Sungtae An,
  Safoora Fatima, Aline Herlopian, Ioannis Karakis, Jonathan~J. Halford,
  Marcus~C. Ng, Emily~L. Johnson, Brian~L. Appavu, Rani~A. Sarkis,
  Gamaleldin~M. Osman, Peter~W. Kaplan, Monica~B Dhakar, Lakshman~Arcot
  Jayagopal, Zubeda~B. Sheikh, Olga Taraschenko, Sarah~E. Schmitt, Hiba~A.
  Haider, Jennifer~A. Kim, Christa~B. Swisher, Nicolas Gaspard,
  Mackenzie~Carpenter Cervenka, Andres A~Rodriguez Ruiz, Jong~Woo Lee, Mohammad
  Tabaeizadeh, Emily~J. Gilmore, Kristen~J. Nordstrom, Ji~Yeoun Yoo, Manisha~G
  Holmes, Susan~T. Herman, Jennifer~A. Williams, Jay~S. Pathmanathan,
  F{\'a}bio~A. Nascimento, Ziwei Fan, Samaneh Nasiri, Mouhsin~M. Shafi,
  Sydney~S. Cash, Daniel~B. Hoch, Andrew~J. Cole, Eric~S. Rosenthal, Sahar~F.
  Zafar, Jimeng Sun, and Michael~Brandon Westover.
\newblock Interrater reliability of expert electroencephalographers identifying
  seizures and rhythmic and periodic patterns in eegs.
\newblock \emph{Neurology}, 100:\penalty0 e1737 -- e1749, 2022.
\newblock URL \url{https://api.semanticscholar.org/CorpusID:254180050}.

\bibitem[Jing et~al.(2020)Jing, Sun, Kim, Herlopian, Karakis, Ng, Halford,
  Maus, Chan, Dolatshahi, Muniz, Chu, Sacca, Pathmanathan, Ge, Dauwels, Lam,
  Cole, Cash, and Westover]{Jing2020DevelopmentOE}
Jin Jing, Haoqi Sun, Jennifer~A. Kim, Aline Herlopian, Ioannis Karakis,
  Marcus~C. Ng, Jonathan~J. Halford, Douglas Maus, Fonda Chan, Marjan
  Dolatshahi, Carlos~F. Muniz, Catherine~J. Chu, Valeria Sacca, Jay~S.
  Pathmanathan, Wendong Ge, Justin Dauwels, Alice~D. Lam, Andrew~J. Cole,
  Sydney~S. Cash, and Michael~Brandon Westover.
\newblock Development of expert-level automated detection of epileptiform
  discharges during electroencephalogram interpretation.
\newblock \emph{JAMA neurology}, 2020.
\newblock URL \url{https://api.semanticscholar.org/CorpusID:204812691}.

\bibitem[Johansen et~al.(2016)Johansen, Jin, Maszczyk, Dauwels, Cash, and
  Westover]{SpikeViaCNN}
Alexander~Rosenberg Johansen, Jing Jin, Tomasz Maszczyk, Justin Dauwels,
  Sydney~S. Cash, and M.~Brandon Westover.
\newblock Epileptiform spike detection via convolutional neural networks.
\newblock In \emph{2016 IEEE International Conference on Acoustics, Speech and
  Signal Processing (ICASSP)}, pages 754--758, 2016.
\newblock \doi{10.1109/ICASSP.2016.7471776}.

\bibitem[Louren{\c{c}}o et~al.(2020)Louren{\c{c}}o, Tjepkema-Cloostermans,
  Teixeira, and van Putten]{ScalpEEG}
Catarina Louren{\c{c}}o, Marleen~C. Tjepkema-Cloostermans, Lu{\'i}s~F.
  Teixeira, and Michel J. A.~M. van Putten.
\newblock Deep learning for interictal epileptiform discharge detection from
  scalp eeg recordings.
\newblock In Jorge Henriques, Nuno Neves, and Paulo de~Carvalho, editors,
  \emph{XV Mediterranean Conference on Medical and Biological Engineering and
  Computing -- MEDICON 2019}, pages 1984--1997, Cham, 2020. Springer
  International Publishing.
\newblock ISBN 978-3-030-31635-8.

\bibitem[Raghu et~al.(2020)Raghu, Sriraam, Temel, Rao, and
  Kubben]{Raghu2020EEGBM}
Shivarudhrappa Raghu, Natarajan Sriraam, Yasin Temel, Shyam~Vasudeva Rao, and
  Pieter~Leonard Kubben.
\newblock Eeg based multi-class seizure type classification using convolutional
  neural network and transfer learning.
\newblock \emph{Neural networks : the official journal of the International
  Neural Network Society}, 124:\penalty0 202--212, 2020.
\newblock URL \url{https://api.semanticscholar.org/CorpusID:211035040}.

\bibitem[Tveit et~al.(2023)Tveit, Aurlien, Plis, Calhoun, Tatum, Schomer,
  Arntsen, Cox, Fahoum, Gallentine, Gardella, Hahn, Husain, Kessler, Kural,
  Nascimento, Tankisi, Ulvin, Wennberg, and Beniczky]{Tveit2023AutomatedIO}
Jesper Tveit, Harald Aurlien, S.~Plis, Vince~D. Calhoun, William~O. Tatum,
  Donald~L. Schomer, Vibeke Arntsen, Fieke~M.E. Cox, Firas Fahoum, William~B.
  Gallentine, Elena Gardella, Cecil~D. Hahn, Aatif~M. Husain, Sudha~Kilaru
  Kessler, Mustafa~Aykut Kural, F{\'a}bio~A. Nascimento, Hatice Tankisi,
  Line~B{\'e}dos Ulvin, Richard~A. Wennberg, and S{\'a}ndor Beniczky.
\newblock Automated interpretation of clinical electroencephalograms using
  artificial intelligence.
\newblock \emph{JAMA Neurology}, 80:\penalty0 805 -- 812, 2023.
\newblock URL \url{https://api.semanticscholar.org/CorpusID:259202197}.

\bibitem[Beede et~al.(2020)Beede, Baylor, Hersch, Iurchenko, Wilcox,
  Ruamviboonsuk, and Vardoulakis]{Beede2020AHE}
Emma Beede, Elizabeth~Elliott Baylor, Fred Hersch, Anna Iurchenko, Lauren
  Wilcox, Paisan Ruamviboonsuk, and Laura~M. Vardoulakis.
\newblock A human-centered evaluation of a deep learning system deployed in
  clinics for the detection of diabetic retinopathy.
\newblock \emph{Proceedings of the 2020 CHI Conference on Human Factors in
  Computing Systems}, 2020.
\newblock URL \url{https://api.semanticscholar.org/CorpusID:213644599}.

\bibitem[Zech et~al.(2018)Zech, Badgeley, Liu, Costa, Titano, and
  Oermann]{Zech2018VariableGP}
John~R. Zech, Marcus~A. Badgeley, Manway Liu, Anthony~Beardsworth Costa,
  Joseph~J. Titano, and Eric~Karl Oermann.
\newblock Variable generalization performance of a deep learning model to
  detect pneumonia in chest radiographs: A cross-sectional study.
\newblock \emph{PLoS Medicine}, 15, 2018.
\newblock URL \url{https://api.semanticscholar.org/CorpusID:49558635}.

\bibitem[Lopes et~al.(2023)Lopes, Cassani, and Falk]{Lopes2023UsingCS}
Marilia Karla~Soares Lopes, Raymundo Cassani, and Tiago~H. Falk.
\newblock Using cnn saliency maps and eeg modulation spectra for improved and
  more interpretable machine learning-based alzheimer's disease diagnosis.
\newblock \emph{Computational Intelligence and Neuroscience}, 2023, 2023.
\newblock URL \url{https://api.semanticscholar.org/CorpusID:256753238}.

\bibitem[Rudin(2018)]{Rudin2018StopEB}
Cynthia Rudin.
\newblock Stop explaining black box machine learning models for high stakes
  decisions and use interpretable models instead.
\newblock \emph{Nature Machine Intelligence}, 1:\penalty0 206--215, 2018.
\newblock URL \url{https://api.semanticscholar.org/CorpusID:182656421}.

\bibitem[Adebayo et~al.(2018)Adebayo, Gilmer, Muelly, Goodfellow, Hardt, and
  Kim]{adebayo2018sanity}
Julius Adebayo, Justin Gilmer, Michael Muelly, Ian Goodfellow, Moritz Hardt,
  and Been Kim.
\newblock Sanity checks for saliency maps.
\newblock \emph{Advances in neural information processing systems}, 31, 2018.

\bibitem[Arun et~al.(2021)Arun, Gaw, Singh, Chang, Aggarwal, Chen, Hoebel,
  Gupta, Patel, Gidwani, Adebayo, Li, and Kalpathy-Cramer]{Arun2021AssessingTT}
Nishanth~Thumbavanam Arun, N.~Gaw, Praveer Singh, Ken Chang, Mehak Aggarwal,
  Bryan Chen, Katharina~Viktoria Hoebel, Sharut Gupta, Jay~B. Patel, Mishka
  Gidwani, Julius Adebayo, Matthew~D Li, and Jayashree Kalpathy-Cramer.
\newblock Assessing the trustworthiness of saliency maps for localizing
  abnormalities in medical imaging.
\newblock \emph{Radiology. Artificial intelligence}, 3 6:\penalty0 e200267,
  2021.
\newblock URL \url{https://api.semanticscholar.org/CorpusID:261967989}.

\bibitem[Rudin et~al.(2021)Rudin, Chen, Chen, Huang, Semenova, and
  Zhong]{Rudin2021InterpretableML}
Cynthia Rudin, Chaofan Chen, Zhi Chen, Haiyang Huang, Lesia Semenova, and Chudi
  Zhong.
\newblock Interpretable machine learning: Fundamental principles and 10 grand
  challenges.
\newblock \emph{ArXiv}, abs/2103.11251, 2021.
\newblock URL \url{https://api.semanticscholar.org/CorpusID:232307437}.

\bibitem[Nascimento et~al.(2022)Nascimento, Barfuss, Jaffe, Westover, and
  Jing]{Nascimento2022AQA}
F{\'a}bio~Augusto Nascimento, Jaden~D. Barfuss, Alex Jaffe, Michael~Brandon
  Westover, and Jin Jing.
\newblock A quantitative approach to evaluating interictal epileptiform
  discharges based on interpretable quantitative criteria.
\newblock \emph{Clinical Neurophysiology}, 146:\penalty0 10--17, 2022.
\newblock URL \url{https://api.semanticscholar.org/CorpusID:253571863}.

\bibitem[Chen et~al.(2018)Chen, Li, Barnett, Su, and Rudin]{Chen2018ThisLL}
Chaofan Chen, Oscar Li, Alina~Jade Barnett, Jonathan Su, and Cynthia Rudin.
\newblock This looks like that: deep learning for interpretable image
  recognition.
\newblock In \emph{Neural Information Processing Systems}, 2018.
\newblock URL \url{https://api.semanticscholar.org/CorpusID:49482223}.

\bibitem[Donnelly et~al.(2021)Donnelly, Barnett, and
  Chen]{deformable_protopnet}
Jon Donnelly, Alina~Jade Barnett, and Chaofan Chen.
\newblock Deformable protopnet: An interpretable image classifier using
  deformable prototypes.
\newblock \emph{CoRR}, abs/2111.15000, 2021.
\newblock URL \url{https://arxiv.org/abs/2111.15000}.

\bibitem[Kane et~al.(2017)Kane, Acharya, Beniczky, Caboclo, Finnigan, Kaplan,
  Shibasaki, Pressler, and van Putten]{Kane2017ARG}
Nick~M. Kane, Jayant~N. Acharya, S{\'a}ndor Beniczky, Luis Otavio S.~F.
  Caboclo, Simon Finnigan, Peter~W. Kaplan, Hiroshi Shibasaki, Ronit~M
  Pressler, and Michel~J.A.M. van Putten.
\newblock A revised glossary of terms most commonly used by clinical
  electroencephalographers and updated proposal for the report format of the
  eeg findings. revision 2017.
\newblock \emph{Clinical Neurophysiology Practice}, 2:\penalty0 170 -- 185,
  2017.
\newblock URL \url{https://api.semanticscholar.org/CorpusID:52271646}.

\end{thebibliography}

\appendix

\section{Appendix}

\subsection{ProtoEEGNet Definition}\label{DefinitionAppendix}

\subsubsection{Backbone}\label{BackboneAppendix}

An input image is first fed into a pre-trained CNN, SpikeNet. This CNN has its final predictive layer removed, so it simply acts as a function $f$ that maps from images to a latent space, creating a tensor of useful features from an input image. In the case of ProtoEEGNet, a 1x128x37 input EEG is convolved into a single 128x1x1 tensor. The original image consists of 128 data points across 37 different channels, creating a sort of 128x37 ``image.'' The end result is 128 different and informative features extracted by the CNN for the entire EEG. 

\subsubsection{Prototype Layer}

Each class is assigned a set number of prototypes, and each prototype is randomly initialized on the d-dimensional unit hypersphere to start. Every prototype is of dimension 128x1x1 and represents the entire image, so we calculate a similarity score per prototype for every image. Prototypes are expressed as $p^{(c,l)}$, where c is the class the prototype belongs to and l is what number it is within that class.

\subsubsection{Similarity Score}

The similarity score g for a given prototype and some input vector in the feature space is calculated using cosine similarity between the two vectors, where both vectors are always ensured to be on the unit hypersphere.

\subsubsection{Fully-Connected Layer}

After the similarity score is calculated for a given prototype, it is sent to the final layer, which is fully connected and has no activation function. This final layer does have normalized outputs using the softmax function: $\frac{e^{q_{k}}}{\sum_{k\in K}e^{q_{k}}}$ and thus acts as a multi-class logistic regression. This final output layer is described as the function $h$ and is used to create final ``probabilities'' for each class so that a prediction can be made based on whichever logit is highest.

\subsection{Prototype Training}\label{TrainingAppendix}

Training was completed in 4 hours on a P100 GPU.

\subsubsection{Stochastic Gradient Descent}

There is a preliminary ``warming'' stage that lasts for 5 epochs, using Stochastic Gradient Descent (SGD). In this stage, only the prototypes are trained, and the weights in both the CNN backbone and the final output layer are frozen. This essentially just allows us to get a head-start on training the prototypes. The CNN is pre-trained, so most of its weights are fixed for now, except the final output weights, which are chosen in a specific way to enable us to train the prototypes effectively; for an output logit of class $k$, any prototype also of class $k$ will have a weight = 1. For any prototype not of class $k$, the assigned weight will be -0.5. This is a simple and consistent configuration that makes training the prototypes while freezing the final layer plausible, as it rewards activation within the correct class and penalizes activation outside the correct class.

After 5 epochs, the CNN weights are unfrozen and are allowed to move with the prototypes. This is done to allow the latent space of the CNN to change and learn as needed to fit the prototypes optimally. This first step may be repeated multiple times, i.e., multiple epochs may be allowed to pass of just doing SGD. The number of epochs is a hyperparameter that can be set.

The total loss function used during SGD, including all regularization terms, can be expressed as follows:
$$l = \textrm{CE}+\lambda_1 l_{\textrm{sep}}+\lambda_2 l_{\textrm{clst}}+\lambda_3 l_{\textrm{ortho}}$$
Where CE is simply the usual cross-entropy. The loss terms are explained in their own section further below.

\subsubsection{Prototype Projection}

In this stage, prototypes are ``pushed'' or projected onto the nearest latent patch of the same class discovered in training as judged by the cosine similarity; i.e., a prototype is pushed onto whatever example is most similar to it. This is useful for two reasons: Firstly, it enables us to explicitly connect the prototypes to an actual example found in the training data. Secondly, it ensures that the prototypes are not in some area of the latent space completely disconnected from the training data. By forcing it to correspond to a training example, a prototype is able to stay in the correct ``neighborhood.'' This step is only applied a few times during training. For example, in the Original ProtoPNet paper \cite{Chen2018ThisLL}, this projection step was only performed at epochs 10, 20, and 30, meaning that between those times, the prototypes were allowed to move freely in space without being explicitly tied to any particular training patch. The epochs at which this step is performed are also hyperparameters.

\subsubsection{Convex Optimization}

In this stage, we freeze the prototypes and the CNN weights and only optimize the final fully connected layer. Because there is no activation function and all other weights are frozen, we are able to use convex optimization rather than gradient descent to optimize this final layer. This allows us to optimize the preset weights used in the previous examples. We also try to force the majority of weights where the class of the prototype does not match the output logit to be 0 (it was previously fixed at -0.5), making the weights sparse. This is because we want the model to have an affirmative reasoning process as opposed to a negative one, where the model chooses a class due to the presence of distinguishing features, rather than a process of elimination where the model picks a class because it is not some other class. Note that this step is always performed directly after the prototype projection stage.

 The problem solved is described below:
$$\min_{w_{h}}\frac{1}{n}\sum_{i=1}^{n}CrsEnt(f\circ g_{p}\circ h(x_{i}),y_{i})+\lambda\sum_{k=1}^{K}\sum_{j:p_{j}\notin P_{k}}|w_{h}^{(k,j)}|$$
where $h$ is the CNN backbone, $g_p$ represents the prototype layers, $f$ represents the final connected layer, and $w_h^{(k,j)}$ is the weight associated in the final connected layer where $k$ is the class in question and $j$ is the class of the prototype being examined. Note that there is an $L_1$ regularization term also being computed here to keep the weights sparse.

\subsubsection{ProtoEEGNet Loss Terms}
On top of the loss function defined above, we have additional regularization terms for clustering, separation, and orthogonality, which are $l_{\textrm{clst}}$, $l_{\textrm{sep}}$, and $l_{\textrm{ortho}}$ respectively.

The clustering loss is expressed as $l_{\textrm{clst}}= -\frac{1}{N}\sum_{i=1}^{N} \max_{\hat{p}^{(c,l)}:c = y^{(i)}}g(\hat{z}^{(i)})^{(c,l)}$. This loss serves to ensure that the latent representation appears similar to at least one prototype, ensuring that prototypes are able to "cluster" around examples that look similar to them. 

The separation term is expressed as $l_{\textrm{sep}}=\frac{1}{N}\sum_{i=1}^{N} \max_{\hat{p}^{(c,l)}:c \neq y^{(i)}}g_p(\hat{z}^{(i)})^{(c,l)}$. This loss serves to reinforce that the similarity between an image and a prototype not of the same class as that image is as small as possible, ensuring that prototypes of different classes are kept separate from one other and thus each class gets its own area of the latent space. 

Finally, the orthogonality loss is expressed as $l_{\textrm{ortho}}=\sum_{c}\|P^{(c)} {P^{(c)}}^T -I^{(L)}\|_{F}^{2}$, where c is a class and $L$ is the number of prototypes within that class, $I$ is the identity matrix in $\mathbb{R}^{L\times L}$, and we use the Frobenius norm. This ensures that prototypes of the same class are as orthogonal to each other as possible, allowing them to own their own portion of the latent space and thus preventing prototypes from being too similar and not representing the full complexity of the class.

\subsection{Data}

\subsubsection{Data Source}

Our dataset includes samples from SpikeNet's MGH dataset \cite{Jing2020DevelopmentOE} in addition to more samples collected from MGH EEG archives. 

\subsubsection{Voter Disagreement}

\begin{figure}[htp]
    \centering
    \includegraphics[width=0.6\linewidth]{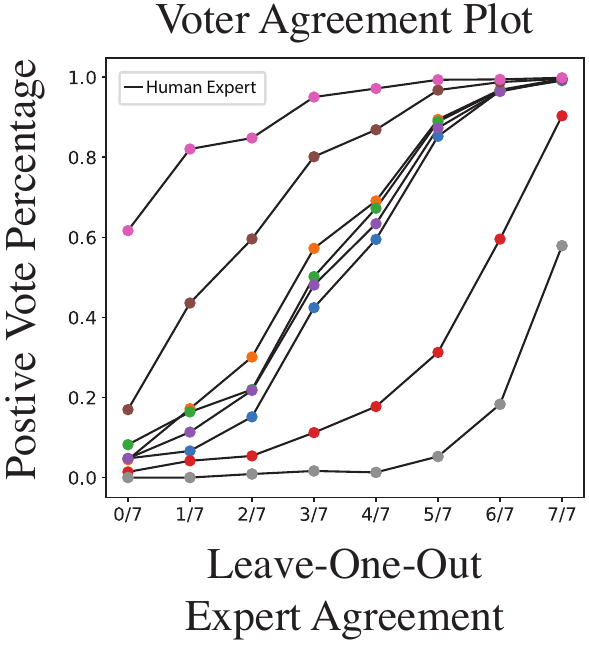}

    \caption{Each line indicates the voting tendencies of a different voter. Given all other experts’ votes (x-axis) for an EEG, each point represents the probability an individual expert will vote that the EEG contains an IED (y-axis).}
    \label{fig:expert_disagreement}
    
\end{figure}

\newpage

\subsection{AUROC Confidence Intervals}\label{AUROCCI}
The 95\% confidence interval on unfiltered data for SpikeNet was (0.840, 0.868) and for ProtoEEGNet -- (0.847, 0.875); on filtered data, for SpikeNet -- (0.902,0.928) and for ProtoEEGNet -- (0.916, 0.940). Confidence intervals were computed via 10,000 rounds of bootstrapping.

\subsection{EEG Class Examples}\label{ClassExamples}
We visualized the different classes of votes below in Figure \ref{fig:expert_votes}. 
\begin{figure}[htp]
    \centering
    \includegraphics[width=1\linewidth]{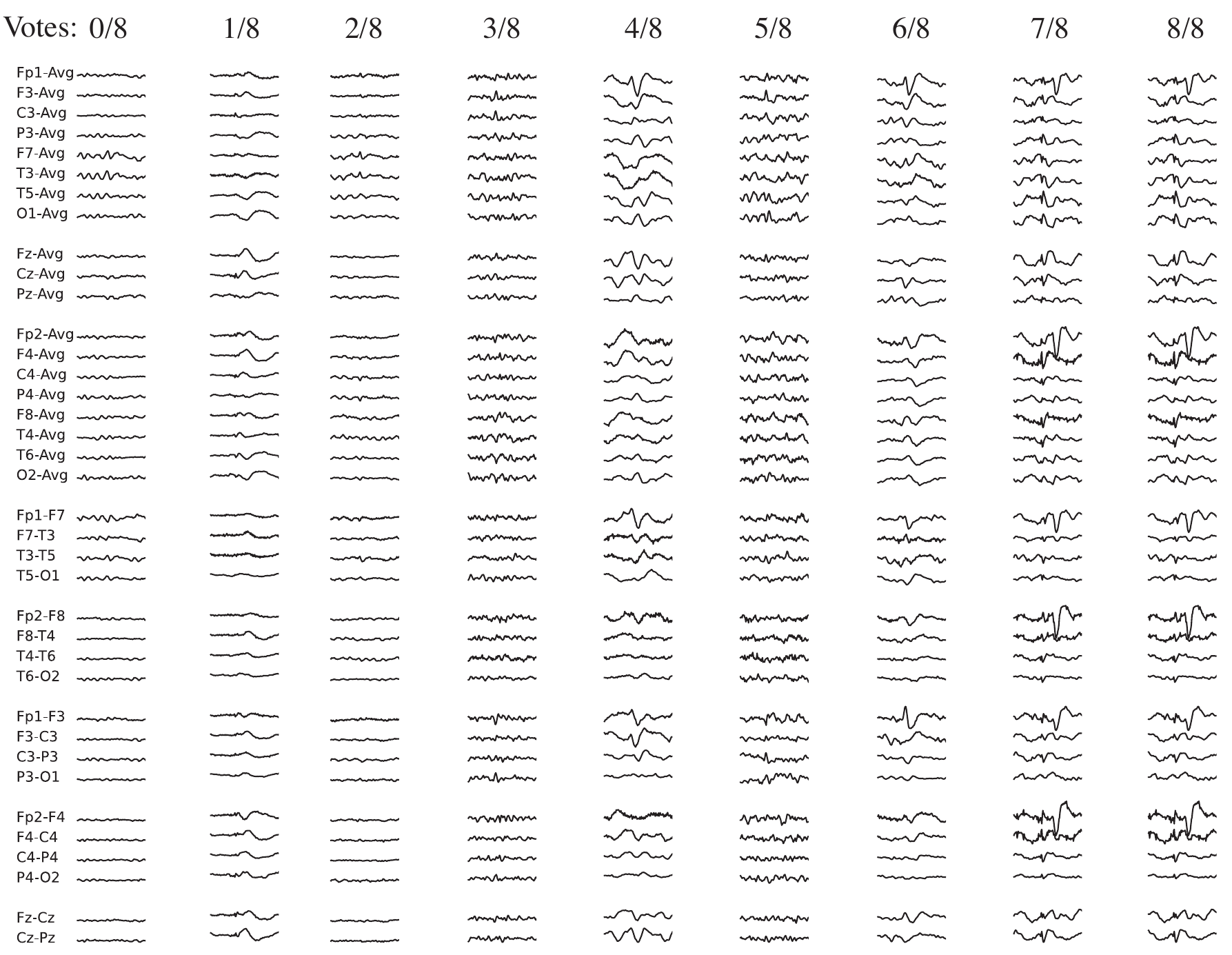}

    \caption{1-second, EEGs arranged in a standard banana bipolar montage. Channels are labeled on the left. Each sample is displayed with the proportion of experts out of 8 that believe the sample contains an IED spike.}
    \label{fig:expert_votes}
    
\end{figure}

\newpage

\newpage

\subsection{Experiment Hyperparameters}\label{TrainingAppendix}

To promote the reproducibility of our technique, we release the hyperparameters used. 
\def\arraystretch{1.8}
\begin{longtable}{|>{\raggedright}p{7.5cm}|>{\raggedright}p{0.8cm}|>{\raggedright}p{4cm}|}\hline
Hyperparameter & Value & Description \tabularnewline\hline
warm\_optimizer\_lrs["prototype\_vectors"] & 0.003 & \multicolumn{1}{m{4cm}|}{Learning rate for the prototype vectors during the warming stage.} \tabularnewline\hline
prototype\_shape & [128, 1, 1] & \multicolumn{1}{m{4cm}|}{Shape of prototypes tensor: total number x number of features x width x height.} \tabularnewline\hline
num\_warm\_epochs & 10 & \multicolumn{1}{m{4cm}|}{Number of warming epochs to be done before joint training.} \tabularnewline\hline
joint\_optimizer\_lrs["joint\_last\_layer\_lr"] & 1.0e-05 & \multicolumn{1}{m{4cm}|}{Learning rate for the last layer during joint training.} \tabularnewline\hline
joint\_optimizer\_lrs["prototype\_vectors"] & 0.05 & \multicolumn{1}{m{4cm}|}{Learning rate for the prototype vectors during joint training.} \tabularnewline\hline
joint\_optimizer\_lrs["conv\_offset"] & 0.003 & \multicolumn{1}{m{4cm}|}{Learning rate for the convolutional offset layers during joint training.} \tabularnewline\hline
joint\_optimizer\_lrs["features"] & 0.001 & \multicolumn{1}{m{4cm}|}{Learning rate for the features during joint training.} \tabularnewline\hline
joint\_optimizer\_lrs["add\_on\_layers"] & 0.001 & \multicolumn{1}{m{4cm}|}{Learning rate for the add-on layers during joining training.} \tabularnewline\hline
prototype\_activation\_function & log &\multicolumn{1}{m{4cm}|}{The activation function associated with prototype layers.} \tabularnewline\hline
push\_start & 70 & \multicolumn{1}{m{4cm}|}{The first epoch at which you push the prototype vectors. Should be > num\_warm\_epochs.} \tabularnewline\hline
warm\_pre\_offset\_optimizer\_lrs["prototype\_vectors"] & 0.003 & \multicolumn{1}{m{4cm}|}{The learning rate for the prototype vectors during the second warming stage.} \tabularnewline\hline
warm\_pre\_offset\_optimizer\_lrs["add\_on\_layers"] & 0.003  & \multicolumn{1}{m{4cm}|}{The learning rate for the add-on layers during the second warming stage.} \tabularnewline\hline
warm\_pre\_offset\_optimizer\_lrs["features"] & 0.001 & \multicolumn{1}{m{4cm}|}{The learning rate for the features during the second warming stage.} \tabularnewline\hline
push\_epochs & [110, 120, 130] & \multicolumn{1}{m{4cm}|}{The list of epochs in which you should push prototypes.} \tabularnewline\hline
joint\_lr\_step\_size & 30 & \multicolumn{1}{m{4cm}|}{The step size for the joint layer optimizer.} \tabularnewline\hline
num\_classes: & 9 & \multicolumn{1}{m{4cm}|}{The number of classes you are classifying between.} \tabularnewline\hline
num\_train\_epochs & 130 & \multicolumn{1}{m{4cm}|}{The number of total train epochs. Note that you should always push on the last epoch, and it is 0-indexed.} \tabularnewline\hline
train\_push\_batch\_size & 75 & \multicolumn{1}{m{4cm}|}{The batch size for the train push step.} \tabularnewline\hline
num\_secondary\_warm\_epochs & 10 & \multicolumn{1}{m{4cm}|}{The number of epochs dedicated to the second warm stage.} \tabularnewline\hline
coefs["clst"] & -0.1 & \multicolumn{1}{m{4cm}|}{The coefficient for the clustering loss.} \tabularnewline\hline
coefs["offset\_weight\_l2:"] & 0.1 & \multicolumn{1}{m{4cm}|}{The coefficient for the offset distance loss.} \tabularnewline\hline
coefs["sep"] & 0 & \multicolumn{1}{|m{4cm}|}{The coefficient for the separation loss.} \tabularnewline\hline
coefs["orthogonality\_loss"] & 0.5 & \multicolumn{1}{m{4cm}|}{The coefficient for the orthogonality loss.} \tabularnewline\hline
coefs["offset\_bias\_l2"] & 1 & \multicolumn{1}{m{4cm}|}{The coefficient for the offset distance loss.} \tabularnewline\hline
coefs["l1"] & 0.01 & \multicolumn{1}{m{4cm}|}{The coefficient for L1 regularization loss on the last layer.} \tabularnewline\hline
coefs["crs\_ent"] & 1.25 &  \multicolumn{1}{m{4cm}|}{The coefficient for the cross-entropy loss.} \tabularnewline\hline
\end{longtable}

\end{document}